\documentclass{article}
\usepackage[left=0.75in, right=0.5in, top=0.5in, bottom=0.5in]{geometry}
\usepackage{cite,graphicx,subfigure,setspace}
\usepackage[small,bf,up]{caption}

\begin{document}
\title{Photonic Crystal Cavities in Silicon Dioxide}
\author{Yiyang Gong* and Jelena Vu\v{c}kovi\'{c} \\
    \small\textit{Department of Electrical Engineering, Stanford University, Stanford, CA 94305} \\
    \small\textit{*email:yiyangg@stanford.edu}}
\twocolumn[
\begin{@twocolumnfalse}
\maketitle
\begin{abstract}
One dimensional nano-beam photonic crystal cavities fabricated in silicon dioxide are considered in both simulation and experiment. Quality factors of over $10^4$ are found via simulation, while quality factors of over $5\times 10^3$ are found in experiment, for cavities with mode volumes of $2.0(\lambda/n)^3$ and in the visible wavelength range (600-716nm). The dependences of the cavity quality factor and mode volume for different design parameters are also considered.
\end{abstract}
\end{@twocolumnfalse}
]
Silicon complementary metal-oxide-semiconductor \\ (CMOS) compatible materials can be integrated with electronics, and thus can have a large impact in computation, communication, and sensors. One important component of an optical system is a photonic cavity, able to store light and spectrally filter signals. The photonic crystal (PC) cavity is one type of optical cavity that allows for both high quality ($Q$) factors and low mode volumes ($V_{mode}$), having ubiquitous use in cavity quantum electrodynamics (cQED) \cite{Dirk_PCQDcontrol}, low threshold lasers \cite{Hatice_arraylaser}, and optical control \cite{Ilya_cavityshift}. One of the most developed PC systems is the system of two dimensional (2D) waveguides and cavities in a free-standing membrane, relying on total internal reflection for confinement in the third dimension. Silicon is transparent in the near infrared regime, and is heavily used for telecommunication purposes near the wavelength of 1.5$\mu$m. However, silicon absorbs heavily in the visible wavelength range, and would be difficult to employ in light emitting and waveguiding devices at the visible wavelengths.

Silicon dioxide (SiO$_{2}$, or silica), on the other hand, is transparent at the visible wavelengths, and similar to silicon, is a promising material due to its low cost, compatibility with electronics, and established fabrication techniques. The main hurdle in manufacturing photonic crystals in silica is its low index of refraction ($n=1.46$). However, perturbation cavities that modulate the index of refraction in a waveguide system can be made even in low index materials. This approach has been used in creating high-$Q$ cavities in 2D PCs in both silicon \cite{Notomi_PChet,Noda_PChet,Noda_PChet1} and low index materials such as silicon nitride (Si$_{3}$N$_{4}$)\cite{Benson_SiNhet}, which has index $n=2.0$, higher than that of SiO$_{2}$. Whereas a full photonic band gap in a 2D photonic crystal is difficult to achieve in low index materials, one dimensional (1D) nano-beam cavities that employ a gentle confinement technique enable high confinement with a small or incomplete band gap, while relying on total internal reflection in both directions perpendicular to the beam length. Recent developments in 1D nano-beam cavities with ``potential well" designs have achieved the same $Q$-factors in silicon as in 2D photonic crystal cavities with comparable mode volumes \cite{MIT_1D,Marko_Si1D}, while also opening the door for high-$Q$ cavities in Si$_{3}$N$_{4}$ for applications of nano-optomechanical coupled resonators \cite{Painter_1Dmodes} and coupling to active materials \cite{Marko_SiN1D}.

In this letter, we design and fabricate 1D nano-beam cavities in silica. 
\begin{figure}[hbtp]
\centering
\includegraphics[width=3.3in]{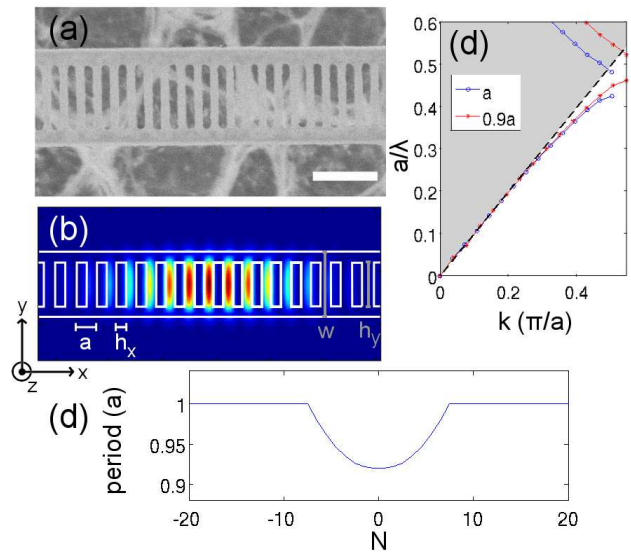}
\caption{(a) The fabricated 1D nano-beam cavity. Marker denotes 1$\mu$m. (b) The electric field intensity ($|E|^2$) of the fundamental mode supported by the cavity. (c) Band diagram for a beam with lattice constant $a$, $w=3a$, $d=0.9a$, $h_{x}=0.5a$, and $h_{y}=0.7w$, and another beam with the same parameters except for lattice constant $a'=0.9a$. The dashed line indicates the light line in free space. (d) Design of the cavity, The plot shows the period ($a$) along the length of the beam as a function of N, the layer number counter from the center of the cavity.}
\label{fig:modes}
\end{figure}
We follow the ladder cavity design used in Si and Si$_{3}$N$_{4}$ \cite{Notomi_1D,Painter_1Dmodes}, but consider a silica slab suspended in free space, with lattice constant $a$, width $w$, slab thickness $d$, hole width $h_{x}$, and hole height $h_{y}$, as shown in Fig. \ref{fig:modes}(b). We first obtain the band diagram of a periodic (or unperturbed) nano-beam waveguide using the three dimensional (3D) finite difference time domain (FDTD) method with Bloch boundary conditions. A sample band diagram is shown in Fig. \ref{fig:modes}(c) for a beam with parameters: lattice constant $a$, $w=3a$, $d=0.9a$, $h_{x}=0.5a$, and $h_{y}=0.7w$, and also for a beam with the same parameters, except with lattice constant $a'=0.9a$. As expected, the structure with the smaller lattice constant has slightly higher band frequencies, as this structure supports modes that have higher overlap with air. Because the lowest band edge mode of the structure with lattice constant $a'$ (mode at the $\pi/a'$ point) lies in the band gap of the structure with lattice constant $a$, it can serve as the defect mode in a beam with lattice constant $a$, which acts as the photonic crystal mirror.

\begin{figure}[hbtp]
\centering
\includegraphics[width=3.3in]{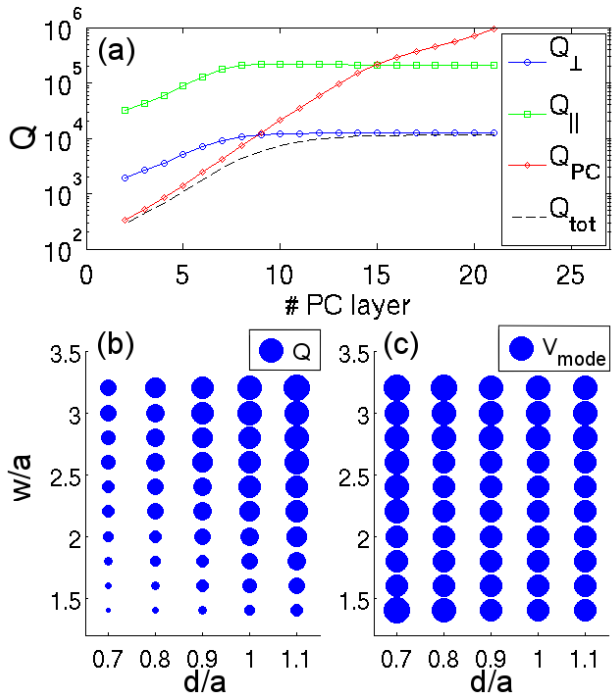}
\caption{(a) The direction specific $Q$-factors of the nano-beam cavity as a function of the number of photonic crystal mirror layers surrounding the cavity. With respect to Fig. \ref{fig:modes}(b), $Q_{\perp}$ corresponds to radiation leaked in the $z$ direction, $Q_{||}$ corresponds to radiation leaked in the $y$ direction, and $Q_{PC}$ corresponds to radiation leaked in the $x$ direction. $Q_{tot}$ is the parallel sum of $Q_{\perp}$, $Q_{||}$, and $Q_{PC}$. (b) The $Q$-factors and (c) the mode volumes of cavities with the same air hole design, but different beam widths and thicknesses. The reference dot sizes are for $Q = 2.0 \times 10^{4}$ and $V_{mode} = 2.0(\lambda/n)^3$ in (b) and (c), respectively.}
\label{fig:fdtd}
\end{figure}

Next, we use the perturbation design suggested by previous references and introduce a parabolic relationship between the lattice constant and the $x$-coordinate, thus forming an optical potential well \cite{Painter_zipper}. In particular, we choose a minimum effective lattice constant of 0.92$a$ at the center of the potential well, and the remainder of the lattice constants as shown in Fig. \ref{fig:modes}(d).  The perturbations of lattice constants span 7 periods away from the center of the cavity. We simulate the full cavity structure again using the FDTD method, and obtain the fundamental TE-like mode with frequency $a/\lambda = 0.454$, $Q=1.6 \times 10^{4}$, $V_{mode} = 2.0 (\lambda/n)^3$, and electric field intensity ($|E|^2$) shown in Fig. \ref{fig:modes}(b). This represents a more than one order of magnitude increase in $Q$-factor and a 7 fold reduction in mode volume compared to a 5$\mu$m diameter silica microdisk cavity with the same silica thickness.

We also simulate the effect of the number of photonic crystal mirror layers on the cavity $Q$. In particular, we define the quality factor in the direction $\hat{i}$ as $Q_{i}=\omega U/P_{i}$, where $\omega$ is the frequency of the mode, $U$ is the total energy of the mode, and $P_{i}$ is the power radiated in the $\hat{i}$ direction. We separate $Q_{tot}=1/(1/Q_{\perp}+1/Q_{||}+1/Q_{PC})$, where $Q_{\perp}$ corresponds to radiation leaked in the $z$ direction in Fig. \ref{fig:modes}(b), $Q_{||}$ corresponds to radiation leaked in the $y$ direction, and $Q_{PC}$ corresponds to radation leaked in the $x$ direction through the photonic crystal mirrors on the silica beam, and plot the results in Fig. \ref{fig:fdtd}(a). We see that the gentle confinement method enables high reflectivity mirrors even in silica, as $Q_{PC}$ continuously increases with the addition of more PC mirror layers. In the case of a 1D nano-beam, $Q$ is limited by loss in the directions where the mode is confined by total internal reflection, namely $Q_{\perp}$ and $Q_{||}$. As seen in Fig. \ref{fig:fdtd}(a), the limiting factor in $Q_{tot}$ is $Q_{\perp}$ in this case. Thus, $Q_{tot}$ could be increased by improving the design of the periods that correspond to the photonic crystal cavity, possibly by parameter search, genetic algorithms \cite{Joel_genetic}, or inverse designs \cite{Dirk_inverse}.

\begin{figure}[hbtp]
\centering
\includegraphics[width=3.3in]{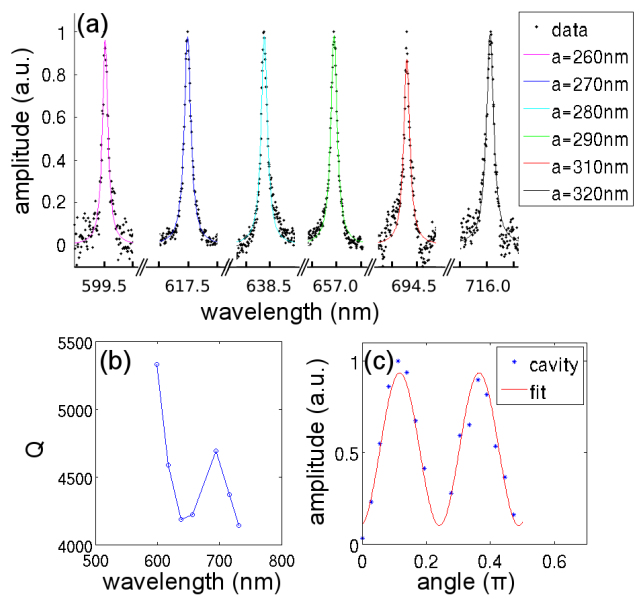}
\caption{(a) The cavities spectra measured in reflectivity from structures with different lattice constants are normalized and shown together, with fits to Lorentizian lineshapes plotted on top of the data points. The cavity spectra are shown from left to right with increasing $a$. (b) The $Q$s of the cavities shown in part (a) plotted against the wavelengths of the cavities. (c) The angle dependence of the reflectivity amplitude. The horizontal axis corresponds to the half-waveplate angle, which is placed in front of the objective lens that is in front of the chip (and thus placed in both the incident and collection paths). The fit to the reflectivity amplitude shows a period very close to $\pi/4$, indicating a linearly polarized cavity mode.}
\label{fig:cavities}
\end{figure}

Furthermore, we simulate the cavity with the same pattern of holes in the $x$-direction, while changing the beam width $w$ and thickness $d$, keeping $h_{x}=0.5a$ and $h_{y}=0.7w$. The resulting limiting $Q$ (the parallel sum of $Q_{\perp}$ and $Q_{||}$) and $V_{mode}$ are shown in Fig. \ref{fig:fdtd}(b) and (c), respectively. We see that the both the $Q$-factor and $V_{mode}$ increase with the slab width and thickness, which is expected as larger cavities have higher confinement but higher mode volumes. In order to employ this cavity in cQED applications, we need to maximize the $Q/V_{mode}$ ratio. Such a maximum is achieved with $w=2.6a$ and $d=1.1a$, with $Q=2.0 \times 10^{4}$ and $V_{mode}=1.8(\lambda/n)^3$.

We next fabricate the cavity with the same dimensions as above. We start by oxidizing a silicon wafer, forming a 270nm layer of SiO$_{2}$ on top of silicon. Next, e-beam lithography is performed with a 250nm layer of ZEP-520A as the resist. After development of the resist, the pattern is transferred to the oxide layer with a CHF$_{3}$:O$_{2}$ chemistry dry etch. Finally, the beam is undercut with a XeF$_{2}$ etcher, removing approximately 4$\mu$m of silicon under the oxide layer. The final fabricated structure is shown in Fig. \ref{fig:modes}(a). We vary the lattice constant $a$ in fabrication to create cavities with a variety of wavelengths.

We characterize the cavities using the cross-polarized reflectivity measurement technique \cite{Hatice_ccavs,Dirk_refl}. In summary, white light linearly polarized 45$^{\circ}$ from the cavity polarization is directed at the cavity through an objective lens from above ($z$ direction), and emission in the polarization orthogonal to the excitation (135$^{\circ}$) is collected also from above and detected by a spectrometer. The fundamental mode of the cavity is linearly polarized in the direction perpendicular to the beam length ($y$-polarized in Fig \ref{fig:modes}(b)), much like the fundamental TE mode of a rectangular slab waveguide. We observe cavities of different resonant wavelengths that span 600-716nm in the visible wavelength range for different lattice constants, as shown in Fig. \ref{fig:cavities}(a). The cavities are shown from left to right with increasing $a$. The measured cavity wavelengths and $Q$s of the cavities are plotted in Fig. \ref{fig:cavities}(b). We also observe a slight decrease in $Q$ with increasing $a$, which is expected from the FDTD simulated trend of decreasing $Q$ with decreasing $d/a$. By placing a half-waveplate in the incident/collection path, we confirm that the observed cavity mode is linearly polarized, as the angle dependent reflectivity amplitude has a period of $\pi/4$ with respect to the half-waveplate angle (Fig. \ref{fig:cavities}(c)).

In conclusion, we have demonstrated a photonic crystal cavity in silica with theoretical $Q$-factors above $10^4$, and experimental $Q$-factors above $5\times 10^3$, while maintaining low mode volumes of 2.0$(\lambda/n)^3$. Such a system could be enhanced by improving the cavity design for higher $Q$-factors through manipulation of the cavity hole placement. In addition, the coupling of emitters to this low index system could be explored. Finally, the silica cavity could be employed to couple optical fields and mechanical oscillations.
 
The authors would like to acknowledge the MARCO Interconnect Focus Center, the Toshiba corporation, and the NSF graduate research fellowship for funding. The authors also acknowledge Maria Makarova for supplying the wafer and Jim Kruger for discussions on fabrication. Fabrication was done at Stanford Nanofabrication Facilities.


\begin{thebibliography}{widest-label}
	\bibitem{Dirk_PCQDcontrol}
		D. Englund, D. Fattal, E. Waks, G. Solomon, B. Zhang, T. Nakaoka, Y. Arakawa, Y. Yamamoto, and J. Vu\v{c}kovi\'{c},  ``Controlling the Spontaneous Emission Rate of Single Quantum Dots in a 2D Photonic Crystal," Phys. Rev. Lett. \textbf{95}, 013904 (2005).
	\bibitem{Hatice_arraylaser}
		H. Altug, D. Englund, and J. Vu\v{c}kovi\'{c}, ``Ultra-fast Photonic Crystal Nanolasers,"  Nature Physics, \textbf{2}, 484 (2006).
	\bibitem{Ilya_cavityshift}
		I. Fushman, E. Waks, D. Englund, N. Stoltz, P. Petroff, and J. Vu\v{c}kovi\'{c}, ``Ultra Fast Nonlinear Optical Tuning of Photonic Crystal Cavities," Appl. Phys. Lett. \textbf{90}, 091118 (2007).
	\bibitem{Notomi_PChet}
		M. Notomi, T. Tanabe, A. Shinya, E. Kuramochi, H. Taniyama, S. Mitsugi, and M. Morita, ``Nonlinear and adiabatic control of high-Q photonic crystal nanocavities," Opt. Expr. \textbf{15}, 17458 (2007).
	\bibitem{Noda_PChet}
		B-S. Song, S. Noda, T. Asano, and Y. Akahane, ``Ultra-high-Q photonic double-heterostructure nanocavity," Nat. Mater. \textbf{4}, 207 (2005).
	\bibitem{Noda_PChet1}
		Y. Takahashi, H. Hagino, Y. Tanaka, B-S. Song, T. Asano, and S. Noda, ``High-Q nanocavity with a 2-ns photon lifetime," Opt. Express \textbf{15}, 17206 (2007).
	\bibitem{Benson_SiNhet}
		M. Barth, N. Nüsse, J. Stingl, B. Löchel, and O. Benson, ``Emission properties of high-Q silicon nitride photonic crystal heterostructure cavities," Appl. Phys. Lett. \textbf{93}, 021112 (2008).

	\bibitem{MIT_1D}
		J. S. Foresi, P. R. Villeneuve, J. Ferrera, E. R. Thoen, G. Steinmeyer, S. Fan, J. D. Joannopoulos, L. C. Kimerling, H. I. Smith, and E. P. Ippen, ``Photonic-bandgap microcavities in optical waveguides, " Nature \textbf{390}, 143 (1997).
	\bibitem{Marko_Si1D}
		P. B. Deotare, M. W. McCutcheon, I. W. Frank, M. Khan, and M. Lon\v{c}ar, ``High Quality factor photonic crystal nanobeam cavities," Appl. Phys. Lett., \textbf{94}, 121106 (2009).
	\bibitem{Painter_1Dmodes}
		M. Eichenfield, R. Camacho, J. Chan, K. J. Vahala, and  O. Painter, ``A picogram- and nanometre-scale photonic-crystal optomechanical cavity," Nature, \textbf{459}, 550 (2009).
	\bibitem{Marko_SiN1D}
		M. W. McCutcheon and M. Lon\v{c}ar, ``Design of an ultrahigh Quality factor silicon nitride photonic crystal nanocavity for coupling to diamond nanocrystals," Optics Express, \textbf{16}, 19136 (2008).
	\bibitem{Notomi_1D}
		M. Notomi, E. Kuramochi, and H. Taniyama, ``Ultrahigh-Q Nanocavity with 1D Photonic Gap," Opt. Expr. \textbf{16}, 11095 (2008).
	\bibitem{Painter_zipper}
		J. Chan, M. Eichenfield, R. Camacho, and O. Painter, ``Optical and mechanical design of a `zipper' photonic crystal optomechanical cavity," Opt. Expr. \textbf{17}, 3802 (2009).
	\bibitem{Joel_genetic}
		J. Goh, I. Fushman, D. Englund, and J. Vu\v{c}kovi\'{c}, ``Genetic Optimization of Photonic Bandgap Structures," Opt. Expr. \textbf{15}, 8218 (2007).
	\bibitem{Dirk_inverse}
		D. Englund, I. Fushman, and J. Vu\v{c}kovi\'{c}, ``General recipe for designing photonic crystal cavities," Opt. Expr. \textbf{13}, 5961 (2005).
	\bibitem{Hatice_ccavs}
		H. Altug and J. Vu\v{c}kovi\'{c}, ``Experimental demonstration of the slow group velocity of light in two-dimensional coupled photonic crystal microcavity arrays," , Appl. Phys. Lett., \textbf{86}, 111102 (2005).
	\bibitem{Dirk_refl}
		D. Englund, A. Faraon, I. Fushman, N. Stoltz, P. Petroff, J. Vu\v{c}kovi\'{c}, ``Controlling Cavity Reflectivity With a Single Quantum Dot," Nature, \textbf{450}, 857 (2007).

\end{thebibliography}
\end{document}